\documentclass[pre,aps,onecolumn,superscriptaddress,nofootinbib]{revtex4-1}
\pdfoutput=1
\usepackage{amsmath, amsthm, amssymb}
\usepackage{amsfonts}
\usepackage{graphicx}
\usepackage{dcolumn}
\usepackage{bm}
\usepackage{textcomp}
\usepackage[normalem]{ulem}

\usepackage[titletoc,title]{appendix}

\usepackage{color}

\usepackage{colordvi}
\usepackage[usenames,dvipsnames]{xcolor}

\usepackage[colorlinks=true, urlcolor=blue, anchorcolor=blue, citecolor=blue,filecolor=blue,linkcolor=blue,menucolor=blue
]{hyperref}

\begin{document}
\title{Geometrical optics of constrained Brownian excursion: from the KPZ scaling to dynamical phase transitions}

\author{Naftali R. Smith}
\email{naftali.smith@mail.huji.ac.il}
\author{Baruch Meerson}
\email{meerson@mail.huji.ac.il}
\affiliation{Racah Institute of Physics, Hebrew University of
Jerusalem, Jerusalem 91904, Israel}

\pacs{05.40.-a, 05.70.Np, 68.35.Ct}

\begin{abstract}
We study a Brownian excursion on the time interval $\left|t\right|\leq T$, conditioned to stay above a moving wall $x_{0}\left(t\right)$ such that $x_0\left(-T\right)=x_0\left(T\right)=0$, and $x_{0}\left(\left|t\right|<T\right)>0$. For a whole class of moving walls, typical fluctuations of the conditioned Brownian excursion are described by the Ferrari-Spohn (FS) distribution and exhibit the Kardar-Parisi-Zhang (KPZ) dynamic scaling exponents $1/3$ and $2/3$. Here we use the optimal fluctuation method (OFM) to study \emph{atypical} fluctuations, which turn out to be quite different. The OFM provides their simple description in terms of optimal paths, or rays, of the Brownian motion. We predict two singularities of the large deviation function, which can be interpreted as dynamical phase transitions, and they are typically of third order.  Transitions of a \emph{fractional} order can also appear depending on the behavior of $x_{0}\left(t\right)$ in a close vicinity of $t=\pm T$.  Although the OFM does not describe typical fluctuations, it faithfully reproduces the near tail of the FS distribution and therefore captures the KPZ scaling. If the wall function $x_{0}\left(t\right)$ is not parabolic near its maximum,  typical fluctuations (which we  probe in the near tail) exhibit a more general scaling behavior with a continuous one-parameter family of scaling exponents.
\end{abstract}

\maketitle
\noindent\large \textbf{Keywords}: \normalsize Large deviations in non-equilibrium systems, Brownian excursion, Dynamical phase transitions.

\tableofcontents
\nopagebreak

\section{Introduction}

Random processes, conditioned to stay away from moving walls, appear in many applications of probability theory and statistical mechanics. We will focus on an important sub-class of these processes: a Brownian excursion  $x\left(t\right)$ with $x \left(-T\right) \! = \! x\left(T\right) \! = \! 0$,  conditioned to stay away from a moving wall $x_0(t)$ such that $x_0(-T)=x_0(T)=0$ and $x_{0}\left(\left|t\right|<T\right)>0$.  This model describes a Brownian particle
which (a) exits the origin at time $t=-T$, (b) returns to the origin at $t=T$, and (c) stays above the moving wall $x_{0}\left(t\right)$ at all $\left|t\right|<T$.

Frachebourg and Martin \cite{Frachebourg2000} encountered this setting when studying the one-dimensional Burgers equation in the inviscid limit with white-noise initial condition, and applying the Hopf-Cole transformation. In this case the effective moving wall is parabolic,
$x_{0}\left(t\right) \! = \! T^2-t^2$. Earlier the parabolic case was studied by Groeneboom \cite{Groeneboom1989}. Ferrari and Spohn \cite{FS} (FS) considered a semicircle $x_{0}\left(t\right) \! = \! \sqrt{T^2-t^2}$.  In both cases (the parabola and the semicircle) one is interested in the statistical properties of $x\left(t\right)-x_{0}\left(t\right)$: the deviations of $x\left(t\right)$ away from the moving wall. FS observed that the semicircle case captures some basic features of the more difficult problem of extreme statistics of $N\gg 1$ non-intersecting  Brownian bridges in one dimension \cite{PrahoferSpohn2002,TW2007,SMCR,Schehr,CH}. Because of the non-intersection, the uppermost Brownian bridge -- an excursion -- typically has a shape of a semicircle. Therefore, as a crude approximation, one can effectively replace all lower-lying Brownian bridges by the single semicircle $x_{0}\left(t\right)$ \cite{FS}. Apart from the semicircle, FS considered a family of more general parabolas $x_{0}\left(t\right)=T^{\gamma}\left(1-t^{2}/T^{2}\right)$
and very briefly discussed general shape functions $x_{0}\left(t\right)=T\,g\left(t/T\right)$.%
\footnote{In the recent years there has been a remarkable progress in the solution of the problem of extreme statistics of non-intersecting Brownian excursions \cite{TW2007,SMCR,Schehr,CH}.  As a result of this progress the semicircular case may have lost some of its initial motivation. However, Brownian excursion, conditioned to stay away from a moving wall, continues to attract attention, and it has been recently generalized in several directions \cite{Ioffe2015,Ioffe2018a,Ioffe2018b,Nechaev,Ioffe2018c}.}

\begin{figure}
\includegraphics[width=0.4\textwidth,clip=]{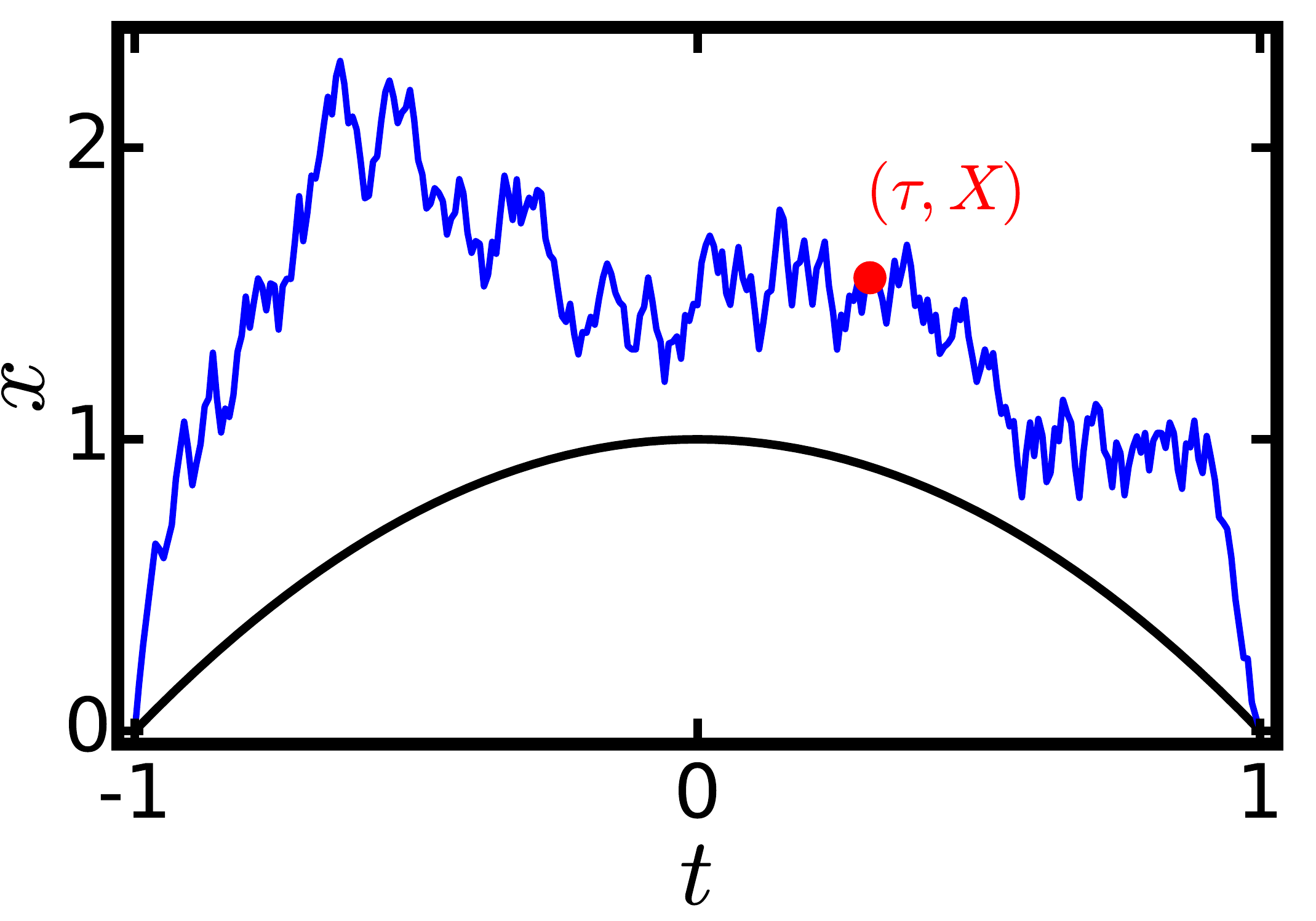}
\caption{A realization of a Brownian excursion which reaches location $X = 1.55$ at time $\tau  = 0.3$ [in rescaled units, see Eq.~(\ref{rescaling})] while avoiding the absorbing wall moving according to $g\left(t\right) = 1-t^2$.}
\label{realization_example}
\end{figure}

Let us introduce the probability distribution $\mathcal{P}\left(X,\tau, T\right)$  of the constrained Brownian excursion $x\left(\tau\right)$ reaching the value $X$ at an intermediate time $\tau \! \in \! \left(-T,T\right)$, see Fig.~\ref{realization_example}. FS obtained the central part of this distribution, which corresponds to \emph{typical} fluctuations of the Brownian excursion away from the moving wall. They found, both for the semicircle and for a parabola $x_0\left(t\right) \! = \! T\left(1-t^{2}/T^{2}\right)$, that typical fluctuations of $X$ scale as $X \! - \! x_0\left(\tau \right) \! \sim \! T^{1/3}$, and that temporal correlations scale as $T^{2/3}$. Somewhat surprisingly, the exponents $1/3$ and $2/3$ coincide with the growth exponent and the correlation exponent, respectively, of the Kardar-Parisi-Zhang (KPZ) equation  \citep{KPZ} which describes an important class of stochastic surface growth. This is in spite of the fact that
the constrained Brownian excursion does \emph{not} belong to the KPZ universality class%
\footnote{Already in their original paper \cite{FS} the authors noticed that finer details of their model differ from those of models of the KPZ universality class. For example, the time correlations
 on the $T^{2/3}$ scale decay exponentially rather than as a power law \cite{FS}. The modern classification identifies the KPZ universality class and its subclasses not only as regards
to scaling exponents, but also as regards to the complete one-point probability distribution (in this case, of $X$) at long times \cite{Corwin,Spohn2016,Dotsenko,Takeuchi2018}. These distributions are also different. Therefore, the FS model shares the scaling exponents with the KPZ class, but does not belong to it.}.

Here we are interested in \emph{atypically} large fluctuations of the constrained Brownian excursion. These fluctuations have not been previously studied. They are described by the \emph{tail} of the distribution $\mathcal{P}\left(X,\tau,T\right)$, where $X-x_0\left( \tau \right)$ is much larger than its typical value. In order to evaluate this tail we will employ
the optimal fluctuation method (OFM), also known as the weak noise theory. In the context of Brownian motion the OFM  is essentially the \emph{geometrical optics} approximation of Brownian motion. Using the OFM, we approximate the probability of observing an unlikely value of $X$ by the probability of the optimal (that is, most likely) path, or ray $x\left(t\right)$, conditioned on reaching the location $X$ at time $\tau$. Mathematically, this approximation corresponds to a saddle-point evaluation of the path integral of the constrained Brownian excursion.

The geometrical optics provides a lucid and instructive insight into the problem by effectively reducing it to an elementary geometric construction. As we show, the optimal path $x\left(t\right)$ is composed of straight-line segments and segments which go along the wall very close to it, $x\left(t\right) \simeq x_0\left(t\right)$.
Further, the geometrical optics reveals \emph{critical lines} -- straight lines in the $\tau, X$ plane, where the number of the segments changes. These lines are the boundaries of complete and partial space-time ``shadows", see Fig.~\ref{parabola_domain}.  Their presence leads to \emph{singularities} in the large deviation function which describes (the logarithm of) $\mathcal{P}\left(X,\tau,T\right)$ at long times. Singularities of large deviation functions are often interpreted as ``dynamical phase transitions", and we adopt this terminology here.
The constrained Brownian excursion is a remarkably simple model, and yet it can exhibit dynamical phase transitions (DPTs) of different orders, which depend on some local properties of the wall function $x_0\left(t\right)$ that we identify.  The physical mechanism behind these transitions -- the space-time shadows -- is markedly different from mechanisms of DPTs in other systems.

The remainder of this paper is organized as follows. In Sec.~\ref{ConstrainedBrownianExcursion} we recap the model and present its OFM formulation. In Sec.~\ref{sec:convex} we use the OFM to calculate $\mathcal{P}\left(X,\tau,T\right)$ for a generic convex upward wall function $x_0\left(t\right)$, and also consider several particular cases of convex upward walls. The non-convex case is briefly discussed in Sec. \ref{sec:nonconvex}. Our main results are summarized and discussed
in Sec.~\ref{disc}. A detailed discussion of the semicircle case, and a comparison of our results with those of FS \cite{FS}, are relegated to the Appendix.

\section{Constrained Brownian excursion and geometrical optics}
\label{ConstrainedBrownianExcursion}

The Brownian motion $x = x\left(t\right)$ can be described by the Langevin equation
\begin{equation}
\label{eq:langevin}
\frac{dx}{dt}=\xi\left(t\right),
\end{equation}
where $\xi$ is a delta-correlated Gaussian noise with zero mean and magnitude $2D$:
\begin{equation}\label{whitenoise}
\left\langle \xi\left(t_{1}\right)\xi\left(t_{2}\right)\right\rangle =2D \, \delta\left(t_{1}-t_{2}\right).
\end{equation}
We consider a Brownian excursion which starts from $x=0$ at $t=-T$ and returns
to $x=0$ for the first time at $t=T$. The excursion is conditioned on escaping absorption by a wall
moving according to the equation
\begin{equation}
x_{0}\left(t\right) = C T^{\gamma} g\left(t/T\right) ,
\end{equation}
such that $g\left(\pm1\right)=0$, $g\left(0\right)=1$ and $\gamma > 0$. $C>0$ is a constant with dimensions length/time$^{\gamma}$.
A realization of this process for the particular case $g\left(t\right) = 1-t^2$ is plotted in Fig.~\ref{realization_example}.

What is the probability density $\mathcal{P}\left(X,\tau,T\right)$ of the conditioned Brownian excursion reaching a point $X$ at time $\tau \! \in \! \left(-T,T\right)$? Clearly $\mathcal{P}$ is nonzero only if $X \! > \! x_0\left(\tau\right)$. We will evaluate $\mathcal{P}$ by using the OFM (or geometrical optics approximation). This approximation (also known as weak noise theory, WKB theory, etc.) can be implemented in several ways. In one of them the WKB ansatz can be applied to the diffusion equation which describes the evolution of the probability density of the position of the Brownian particle \cite{FW}. Here we  will use a more direct approach. Starting from Eqs.~(\ref{eq:langevin}) and~(\ref{whitenoise}), one can express the unconstrained path probability of the Brownian excursion as a path integral $\propto\text{exp}\left(-S\right)$, where
\begin{equation}\label{Action}
S=\frac{1}{4D}\int_{-T}^T \left(\frac{dx}{dt}\right)^2\,dt,
\end{equation}
see\textit{ e.g.} Ref.  \cite{legacy}. The conditional probability distribution $\mathcal{P}\left(X,\tau,T\right)$ is given by the ratio of the probabilities of a Brownian excursion with and without the additional constraint $x\left(\tau\right)=X$. Each of these two probabilities is given by a path integral over all possible paths. The OFM assumes that each of the path integrals is dominated by the action along a single ``optimal" path, or ray, $x\left(t\right)$, for which the action~(\ref{Action}) is minimum.
This observation, combined with a simple rescaling of variables, brings  important implications. Indeed,
let us rescale the coordinate and time as follows:
\begin{equation}\label{rescaling}
 \frac{t}{T} \to t,\quad \text{and}\quad \frac{x}{CT^{\gamma}} \to x .
\end{equation}
Correspondingly, the intermediate time $\tau$ is rescaled by $T$,  $X$ is rescaled by $CT^{\gamma}$, and the rescaled wall function is simply $g\left(t\right)$.
As a result, the distribution, as predicted by the OFM and Eq.~(\ref{Action}), has the following scaling form:
\begin{equation}\label{action1}
-\ln\mathcal{P}\left(X,\tau,T\right)\simeq \frac{ C^2T^{2\gamma-1}}{D}\, s\left(\frac{X}{CT^{\gamma}},\frac{\tau}{T}\right) .
\end{equation}
The large deviation function $s\left(\dots\right)$ is given by $s=s_{\text{c}}-s_{\text{u}}$ where $s_{\text{c}}$ and $s_{\text{u}}$ are the rescaled actions
\begin{equation}
\label{action2}
\frac{1}{4}\int_{-1}^1 \left(\frac{dx}{dt}\right)^2\,dt\,,
\end{equation}
evaluated over the optimal constrained and unconstrained optimal paths $x_{\text{c}}\left(t\right)$ and $x_{\text{u}}\left(t\right)$, respectively. Here the terms ``constrained'' and ``unconstrained'' refer only to the constraint $x\left(\tau\right) =  X$, and they are the origin of our notations `c' and `u' (respectively) for the subscripts.

Equation~(\ref{action1}) has  two immediate implications. First, for $\gamma=1$ the large-deviation scaling form~(\ref{action1}) is, in general, different from the KPZ scaling $X-x_0\left(T\right) \sim T^{1/3}$, observed for typical fluctuations \cite{FS}. As we will see shortly,
there is a joint region (that we call the \emph{near tail}) where the two scalings coincide. Secondly, Eq.~(\ref{action1})  implies that, for $\gamma>1/2$, the OFM becomes asymptotically exact in the limit $T \to \infty$, as long as the function $s$ is not too small.  The latter condition requires that the deviations from the wall, $X-x_{0}\left(t\right)$, be much larger than the typical fluctuations (which, for $\gamma=1$, exhibit the KPZ scaling $\sim T^{1/3}$ \cite{FS}).

\section{Optimal path for convex upward $x_{0}\left(t\right)$}
\label{sec:convex}

The optimal path of the constrained Brownian excursion must minimize the action~(\ref{action2}) under the condition that the path stays above the wall $x\left(t\right) \ge g\left(t\right)$. This problem of one-sided variations is a standard problem of the variational calculus, see \textit{ e.g.}  Ref.~\cite{Elsgolts}.
Its solution consists of alternating segments of two different types: (i) where $x\left(t\right)  = g\left(t\right)$ and (ii) where $x\left(t\right)$ satisfies the Euler-Lagrange equation $d^{2}x/dt^{2}=0$, that is it is a straight line. At points where two segments meet they must have a common tangent \citep{Elsgolts}.

We will begin by considering the particular case where the location of the wall is described by a parabola,
\begin{equation}
g\left(\frac{t}{T}\right)=1-\left(\frac{t}{T}\right)^{2}.
\end{equation}
This case is exactly OFM-solvable, and it exhibits all of the generic features of the model.
We then generalize some of the results to a generic convex upward wall function $g''\left(t\right) <  0$.
Afterwards, we focus on several somewhat less generic, but still interesting examples:
a family of generalized parabolas:
\begin{equation}\label{parabolas}
g\left(\frac{t}{T},\nu\right)=1-\left|\frac{t}{T}\right|^{\nu},
\end{equation}
where $\nu>1$, the cosine
\begin{equation}
g\left(\frac{t}{T}\right)=\cos\left(\frac{\pi t}{2T}\right)
\end{equation}
and the ``tent''
\begin{equation}	
g\left(\frac{t}{T}\right)=1-\left|\frac{t}{T}\right|.
\end{equation}
A detailed study of the semicircle
\begin{equation}\label{circle}
g\left(\frac{t}{T}\right)=\sqrt{1-\left(\frac{t}{T}\right)^{2}}
\end{equation}
is presented in the Appendix.

\subsection{Parabola}
\label{Sec:parabola_nu2}

\begin{figure}
\includegraphics[width=0.32\textwidth,clip=]{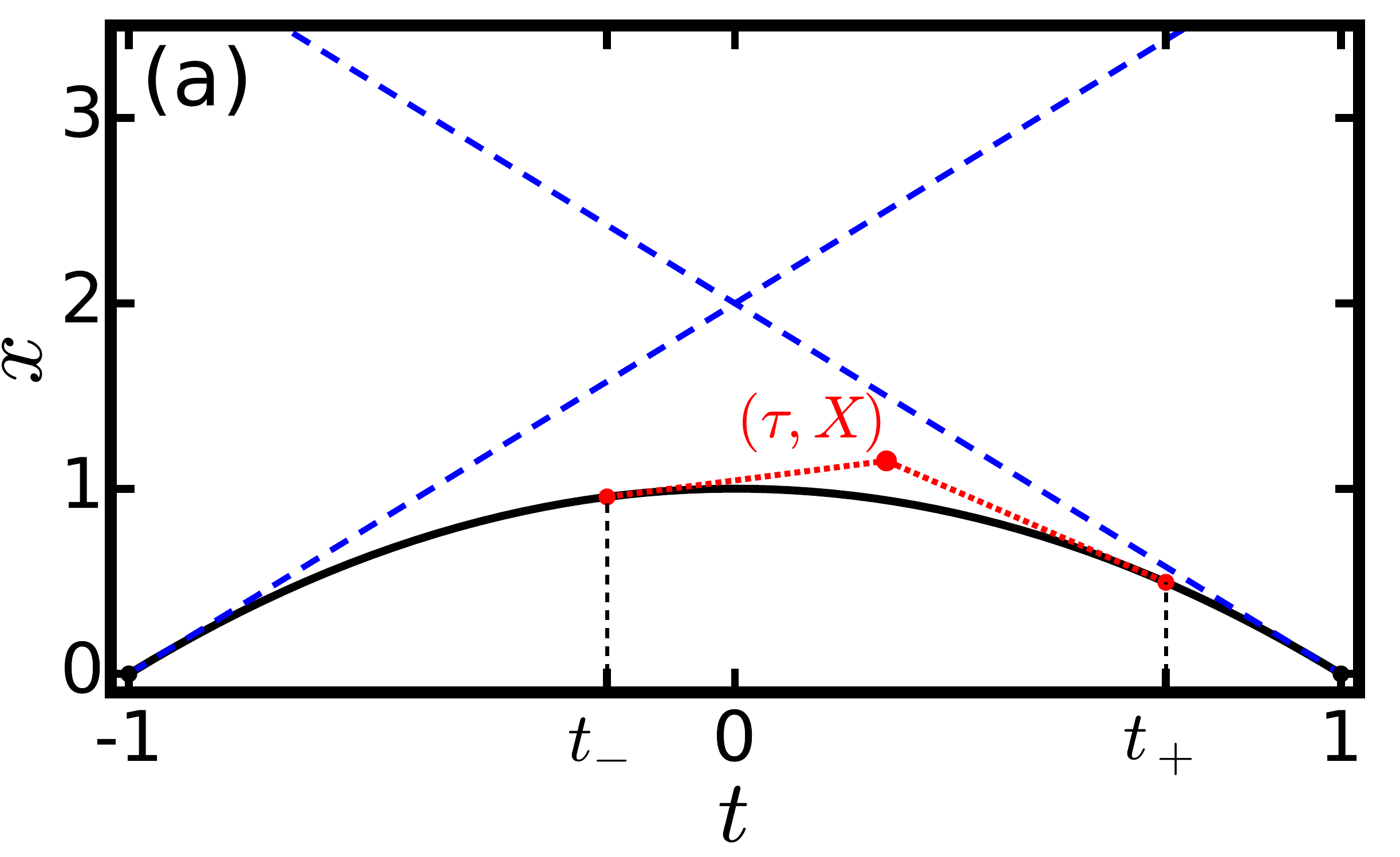}
\includegraphics[width=0.32\textwidth,clip=]{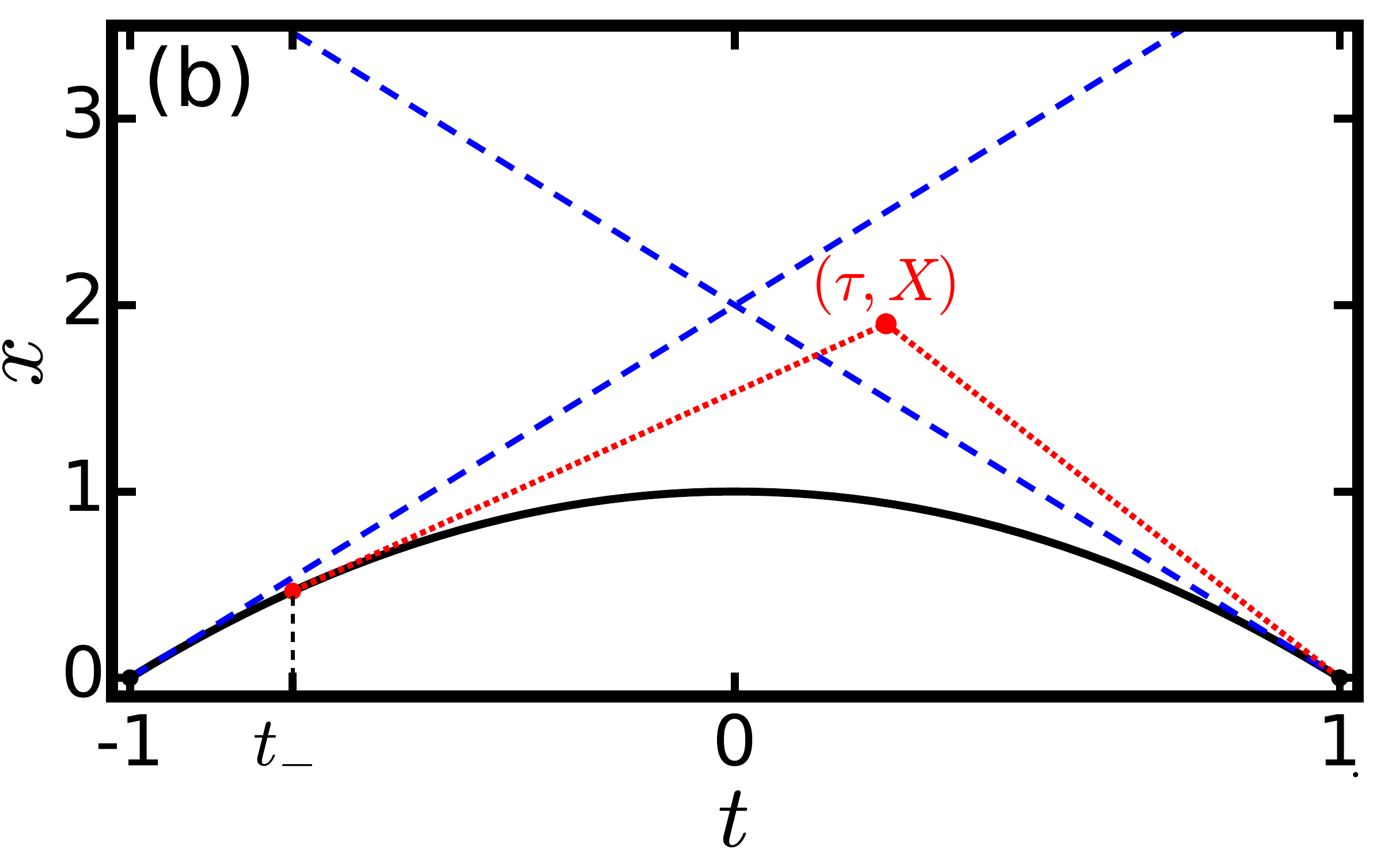}
\includegraphics[width=0.32\textwidth,clip=]{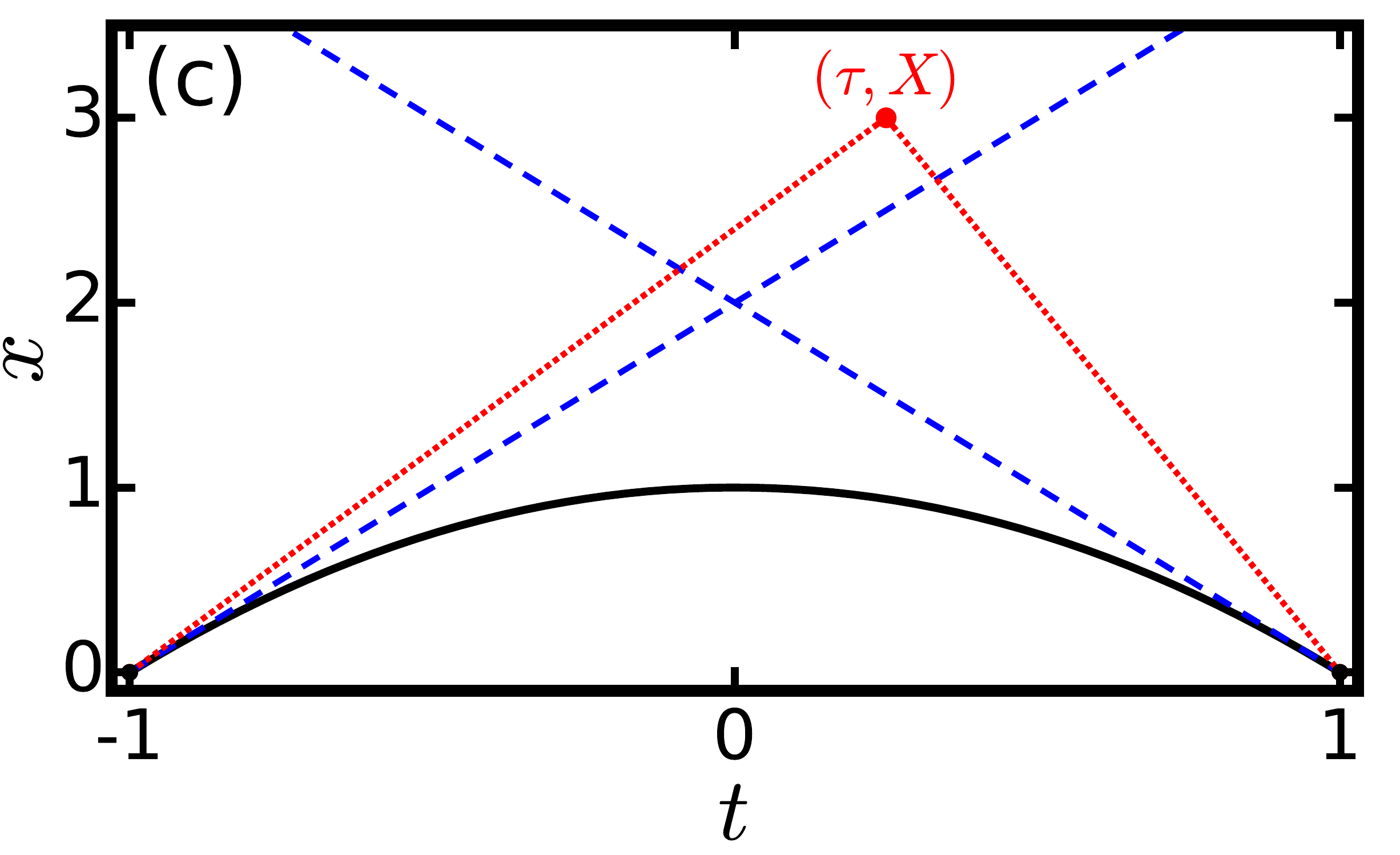}
\caption{Optimal paths for the parabolic wall, $g\left(t\right) = 1-t^2$. Solid line: the wall.
Dotted line: the optimal path constrained on $x\left(t=\tau\right)=X$, at times where $x\left(t\right)\ne g\left(t\right)$, in the subcritical (a), intermediate (b) and supercritical (c) regimes.
The boundaries between the regimes (dashed) are given by the tangents to $g\left(t\right)$ at $t = \pm 1$. Crossing these boundaries  in the vertical direction is accompanied by a third-order dynamical phase transition: a jump in the third derivative $\partial^{3}s/\partial X^{3}$ of the large deviation function $s\left(X,\tau\right)$.}
\label{parabola_domain}
\end{figure}

Here we consider the parabolic wall function $g\left(t\right)=1-t^{2}$  \citep{Groeneboom1989,Frachebourg2000, FS}.
The optimal unconstrained path coincides with the wall's location $x_{\text{u}}\left(t\right)=1-t^2$, and the only nontrivial part of the OFM problem is finding the optimal constrained path $x_{\text{c}}\left(t\right)$. There are three regimes of interest: subcritical, intermediate and supercritical, see Fig.~\ref{parabola_domain}.
In the subcritical regime
$$1-\tau^{2}<X<2\left(1-\left|\tau\right|\right)$$
the optimal constrained path $x_{\text{c}}\left(t\right)$ is obtained via the construction of two tangents from the point $\left(\tau,X\right)$ in the $tx$ plane to the graph of the function $g\left(t\right)$ \citep{Elsgolts}. The left ($t_-$) and right ($t_+$) points of tangency are the solutions to the equation
\begin{equation}
\label{eq:tangent_g_parabola}
X-1+t_{\pm}=-2t_{\pm}\left(\tau-t_{\pm}\right),
\end{equation}
and are given by
\begin{equation}
\label{t_pm_nu2}
t_{\pm}=\tau\pm\sqrt{\tau^{2}-\left(1-X\right)}.
\end{equation}
The optimal constrained path is given in terms of $t_\pm$ by
\begin{equation}
\label{eq:optimal_path_parabola}
x_{\text{c}}\left(t\right)=\begin{cases}
1-t^{2} & t\notin\left[t_{-},t_{+}\right],\\
1-t_{-}^{2}-2t_{-}\left(t-t_{-}\right), & t_{-}\le t\le\tau,\\
X-2t_{+}\left(t-\tau\right), & \tau\le t\le t_{+}.
\end{cases}
\end{equation}
Since $x_{\text{c}}\left(t\right) \neq x_{\text{u}}\left(t\right)$ only at times $t \in  \left[t_{-},t_{+}\right]$, it is sufficient to evaluate the rescaled actions over the interval $\left[t_{-},t_{+}\right]$, that is $s\left(X,\tau\right)=\tilde{s}_{\text{c}}-\tilde{s}_{\text{u}}$ where
\begin{eqnarray}
\tilde{s}_{\text{c}}&=&\frac{1}{4}\int_{t_{-}}^{t_{+}}\left(\frac{dx_{\text{c}}}{dt}\right)^{2}dt=\left(\tau-t_{-}\right)t_{-}^{2}+\left(t_{+}-\tau\right)t_{+}^{2}=2\sqrt{\tau^{2}-1+X}\left(2\tau^{2}-1+X\right),\\\tilde{s}_{\text{u}}&=&\frac{1}{4}\int_{t_{-}}^{t_{+}}\left(\frac{dx_{\text{u}}}{dt}\right)^{2}dt=\frac{1}{4}\left[\left(t_{+}-t_{-}\right)-\frac{1}{3}\left(t_{+}^{3}-t_{-}^{3}\right)\right]=\frac{2\sqrt{\tau^{2}-1+X}\left(4\tau^{2}-1+X\right)}{3}.
\end{eqnarray}
As a result,
\begin{equation}
\label{eq:s_parabola_regime_1}
s\left(X,\tau\right)=\tilde{s}_{\text{c}}-\tilde{s}_{\text{u}}=\frac{4}{3}\left(\tau^{2}+X-1\right)^{3/2},\qquad1-\tau^{2}\le X\le2\left(1-\left|\tau\right|\right).
\end{equation}
This result is valid at $X\leq 2\left(1-\left|\tau\right|\right)$, when the tangency points $t_\pm$ lie within the interval $\left[-1,1\right]$.

In the intermediate regime,
$$2\left(1-\left|\tau\right|\right) \le X \le 2\left(1+\left|\tau\right|\right),$$
the calculation is modified as follows: if $\tau > 0$, we replace $t_+$ by $1$, and if $\tau < 0$, we replace $t_-$ by $-1$.
The result is
\begin{equation}
\label{eq:s_parabola_regime_2}
s\left(X,\tau\right)=-\frac{\left(\!\sqrt{\tau^{2}+X-1}-\left|\tau\right|\right)^{3}\!\!+1}{3}+\frac{1}{4}\left\{ \!\frac{X^{2}}{1-\left|\tau\right|}+\frac{\left[\left(\left|\tau\right|-\sqrt{\tau^{2}+X-1}\right)^{2}\!\!+X\!-1\right]^{2}}{\sqrt{\tau^{2}+X-1}}\right\} \!,\;\,2\left(1-\left|\tau\right|\right)\! \le \!X\! \le \!2\left(1+\left|\tau\right|\right).
\end{equation}

In the supercritical regime $X \ge 2\left(1+\left|\tau\right|\right)$
the optimal constrained path is unaffected by the wall and given by two straight lines:
\begin{equation}
\label{eq:two_straight_lines}
x_{\text{c}}\left(t\right)=\begin{cases}
\frac{1+t}{1+\tau}X, & -1 \le t \le \tau,\\
\frac{1-t}{1-\tau}X, & \tau \le t \le 1.
\end{cases}
\end{equation}
As a result, in this regime
\begin{equation}
\label{eq:s_c_far_tail}
s_{\text{c}}=\frac{X^{2}}{2\left(1-\tau^{2}\right)}.
\end{equation}
It is straightforward to calculate the action along the optimal unconstrained path:
\begin{equation}
s_{\text{u}}=\frac{1}{4}\int_{-1}^{1}\left(\frac{dx_{\text{u}}}{dt}\right)^{2}dt=\int_{-1}^{1}t^{2}dt=\frac{2}{3}
\end{equation}
Altogether, we find
\begin{equation}
\label{eq:s_parabola_regime_3}
s\left(X,\tau\right)=\frac{X^{2}}{2\left(1-\tau^{2}\right)}-\frac{2}{3}, \qquad X\ge2\left(1+\left|\tau\right|\right),
\end{equation}
so that the far tail of the $\mathcal{P}$-distribution is Gaussian, and the wall only contributes the constant $-2/3$.

\begin{figure}
\includegraphics[width=0.4\textwidth,clip=]{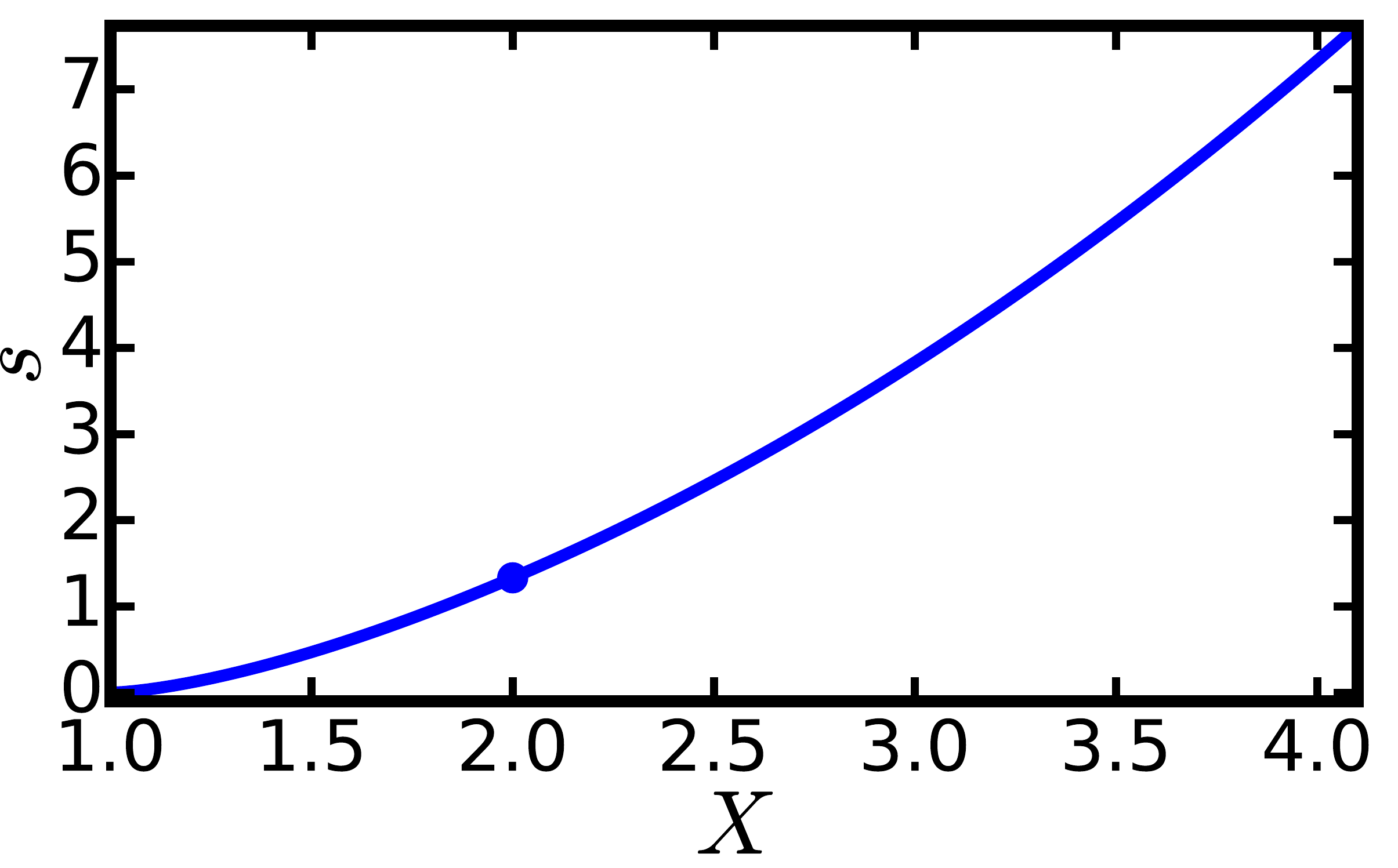}
\caption{The large-deviation function $s\left(X,\tau=0\right)$ for the parabola $g\left(t\right)=1-t^{2}$, see Eqs.~(\ref{eq:s_parabola_regime_1}) and~(\ref{eq:s_parabola_regime_3}). At $X<2$ the scaling $s\sim\left(X-1\right)^{3/2}$ is observed. This scaling breaks down at $X=2$ where a third-order dynamical phase transition occurs, corresponding to a jump in the third derivative $d^{3}s/dX^{3}$. At $X > 2\;$ $s\left(X\right)$ is a quadratic function of $X$, corresponding to a Gaussian tail of the distribution $\mathcal{P}\left(X,\tau,T\right)$.}
\label{s_X_parabola}
\end{figure}

The large deviation function $s\left(X,\tau\right)$, as described by Eqs.~(\ref{eq:s_parabola_regime_1}),~(\ref{eq:s_parabola_regime_2}) and~(\ref{eq:s_parabola_regime_3}), is continuous together with its first and second derivatives, $\partial s/\partial X$ and $\partial^{2}s/\partial X^{2}$, at each of the two transition lines $X=2\left(1\pm\left|\tau\right|\right)$. The third derivative $\partial^{3}s/\partial X^{3}$, however, jumps at the transition lines, which corresponds to a  third-order dynamical phase transition.
In the particular case $\tau = 0$ these two phase transitions merge into one third-order transition at $X=2$, and the intermediate regime disappears.
$s\left(X,\tau = 0\right)$ is plotted in Fig.~\ref{s_X_parabola}.

In analogy with geometrical optics, the supercritical, intermediate and subcritical regimes correspond (respectively) to lit, partially lit and dark areas in the $tx$ plane, if one were to interpret $t$ as a spatial coordinate, and given point light sources at the points $\left(\pm1,0\right)$ and an opaque wall $g\left(t\right)$.

\subsection{General}
\label{sec:convexg_general}

We now extend some of the results of the previous subsection to a generic convex upward wall function $g''\left(t\right)<0$. The extension is fairly straightforward, and the qualitative properties of the system remain mostly unaffected.

In the subcritical regime, the optimal constrained path is given by a similar construction to the one we showed for the parabola.
Eqs.~(\ref{eq:tangent_g_parabola}) and~(\ref{eq:optimal_path_parabola}) give way to
\begin{equation}
\label{eq:tangent_g}
X-g\left(t_{\pm}\right)=\left(\tau-t_{\pm}\right)g'\left(t_{\pm}\right)
\end{equation}
and
\begin{equation}
\label{eq:optimal_path_convex_g}
x_{\text{c}}\left(t\right)=\begin{cases}
g\left(t\right), & t\notin\left[t_{-},t_{+}\right],\\
g\left(t_{-}\right)+g'\left(t_{-}\right)\left(t-t_{-}\right), & t_{-}<t<\tau,\\
X+g'\left(t_{+}\right)\left(t-\tau\right), & \tau<t<t_{+},
\end{cases}
\end{equation}
respectively.
Defining $\tilde{s}_{\text{c}}$ and $\tilde{s}_{\text{u}}$ as we did for the parabola, we find that $s\left(X,\tau\right)=\tilde{s}_{\text{c}}-\tilde{s}_{\text{u}}$ where
\begin{eqnarray}
\label{eq:s_t_g}
\tilde{s}_{\text{c}}&=&\frac{1}{4}\left(\tau-t_{-}\right)\left[g'\left(t_{-}\right)\right]^{2}+\frac{1}{4}\left(t_{+}-\tau\right)\left[g'\left(t_{+}\right)\right]^{2} = \frac{\left[X-g\left(t_{-}\right)\right]^{2}}{4\left(\tau-t_{-}\right)}+\frac{\left[X-g\left(t_{+}\right)\right]^{2}}{4\left(t_{+}-\tau\right)},\\
\label{eq:s_b_g}
\tilde{s}_{\text{u}}&=&\frac{1}{4}\int_{t_{-}}^{t_{+}}\left[g'\left(t\right)\right]^{2}dt.
\end{eqnarray}

In the intermediate regime, where Eq.~(\ref{eq:tangent_g}) admits exactly one solution within the interval $\left[-1,1\right]$, one proceeds by replacing $t_-$ by $-1$ or $t_+$ by $1$, in a similar manner to that described in the previous subsection.

In the supercritical regime, where Eq.~(\ref{eq:tangent_g}) admits no solutions within the interval $\left[-1,1\right]$, the optimal constrained path and its corresponding action are given by Eqs.~(\ref{eq:two_straight_lines}) and~(\ref{eq:s_c_far_tail}) respectively, so that the far tail of the $\mathcal{P}$-distribution is Gaussian. This tail is universal and independent of the wall function $g\left(t\right)$. The wall function only contributes a constant, which arises from $s_{\text{u}}$, and we obtain
\begin{equation}
\label{eq:s_far_tail}
s=s_{\text{c}}-s_{\text{u}}=\frac{X^{2}}{2\left(1-\tau^{2}\right)}-\frac{1}{4}\int_{-1}^{1}\left[g'\left(t\right)\right]^{2}dt.
\end{equation}

The boundaries between the regimes are given by the tangents to $g\left(t\right)$ at $t =  \pm 1$, as we already showed for the parabola, see Fig.~\ref{parabola_domain}.

In the near tail, $\left|X-g\left(\tau\right)\right|\ll g\left(\tau\right)$, we can solve Eq.~(\ref{eq:tangent_g}) by expanding the function $g\left(t\right)$ around $t=\tau$ up to second order%
\footnote{We assume here that that $g''\left(\tau\right)<0$ is finite. We will relax these conditions in sections \ref{Sec:parabola} and \ref{sec:tent}.}, yielding
\begin{equation}
\label{eq:t_pm_g}
t_{\pm}=\tau\pm\sqrt{-2\frac{X-g\left(\tau\right)}{g''\left(\tau\right)}}.
\end{equation}
Plugging Eq.~(\ref{eq:t_pm_g}) into Eqs.~(\ref{eq:s_t_g}) and~(\ref{eq:s_b_g}) and keeping leading-order terms in the expansion of $g\left(t\right)$ around $t=\tau$, we obtain
\begin{equation}
\label{eq:s_typical_fluctuations_convex_g}
s\left(X,\tau\right)\simeq\frac{2\sqrt{2}}{3}\sqrt{-g''\left(\tau\right)}\,\left(\Delta X\right)^{3/2} ,
\end{equation}
where $\Delta X \equiv X-g\left(\tau\right)$. Plugging Eq.~(\ref{eq:s_typical_fluctuations_convex_g}) into~(\ref{action1}), we obtain the near tail of the $\mathcal{P}$-distribution. In the original variables
\begin{equation}
\label{eq:distribution_physical_variables_convex_g}
-\ln \mathcal{P}\simeq\frac{2 \, \sqrt{-2Cg''\left(\frac{\tau}{T}\right)}\,\left[X-x_{0}\left(\tau\right)\right]^{3/2}}{3D\sqrt{T}},
\end{equation}
in full agreement%
\footnote{\label{footnote:FScomment} Note that in Ref.~\cite{FS} $D=1/2$ and $C=1$.}
with the result quoted in Sec. 5(i) of Ref. \cite{FS}.

Ref. \citep{FS} mostly dealt with typical fluctuations away from the semicircle $g\left(t\right)=\sqrt{1-t^{2}}$. We determine the entire large-deviation function for the semicircle in the Appendix. As we show there, the near-tail asymptotic of the large deviation function coincides with the tail of
the FS distribution, up to pre-exponential corrections which are beyond the accuracy of the leading-order OFM.

According to Eq.~(\ref{eq:distribution_physical_variables_convex_g}), the scaling of typical fluctuations is
\begin{equation}
\label{eq:deltaX_T_third}
X-x_0\left(\tau\right)\sim\frac{D^{2/3}T^{1/3}}{C^{1/3}\left[-g''\left(\frac{\tau}{T}\right)\right]^{1/3}}.
\end{equation}
The correlation time $t_c$ can be evaluated by calculating $\tau-t_{-}$ (or equivalently $t_{+} \! -\tau$) for  a typical $X$. This yields
\begin{equation}
\label{eq:t_c}
t_{c}\sim\frac{D^{1/3}T^{2/3}}{C^{2/3}\left[-g''\left(\frac{\tau}{T}\right)\right]^{2/3}}.
\end{equation}
The scalings $X-x_0\left(\tau\right) \sim T^{1/3}$ and $t_c \sim T^{2/3}$ were found by FS~\cite{FS}. Interestingly, the exponents $1/3$ and $2/3$ coincide with $\beta$ and $1/z$ -- the growth and correlation exponents, respectively -- of the KPZ equation \citep{KPZ}.

As the reader may have noticed,  the optimal constrained path $x_{\text{c}}\left(t\right)$ always has a corner singularity at $t = \tau$. This singularity can be better understood by considering an alternative (but equivalent) formulation of the OFM's variational problem, where the constraint $x\left(\tau\right) = X$ is taken into account by adding the integral term
$$\Lambda\int_{-1}^{1}x\left(t\right)\delta\left(t-\tau\right)dt\, \equiv \,\Lambda x\left(\tau\right)$$
(where $\Lambda$ is a Lagrange multiplier) to the action~(\ref{action2}).  The solution $x_{\text{c}}\left(t\right)$ of the ensuing Euler-Lagrange equation,
$$
\frac{d^{2}x\left(t\right)}{dt^{2}}+2\Lambda\delta\left(t-\tau\right)=0,
$$
has a corner singularity at $t=\tau$.

The following three subsections deal, through examples, with somewhat less generic, but still interesting cases.
In subsection \ref{Sec:parabola} we consider a family of generalized parabolas in order to understand how the local properties of the wall function $g\left(t\right)$ near the measurement time affect the scaling behavior of typical fluctuations (that we probe in the near tail).

As mentioned above, the tangents to the function $g\left(t\right)$ at $t = \pm 1$ are phase transition lines for a generic convex upward $g\left(t\right)$, when  $g'\left(t=\pm1\right)$ is finite. In this case  the order of the phase transitions is determined by the local properties of $g\left(t\right)$ near $t=\pm 1$, and it can be different from the ``typical" third order, see Sec.~\ref{sec:fractional_order} below. The crucial role of the points $t=\pm 1$ is in sharp contrast to typical fluctuations, which are determined only by local properties of the wall function $g\left(t\right)$ near the measurement time $t=\tau$, see Eq.~(\ref{eq:distribution_physical_variables_convex_g}).

If the wall function $g\left(t\right)$ has a corner singularity, the scaling of typical fluctuations and the critical behavior are both strongly affected. We show this in Sec.~\ref{sec:tent} by considering the ``tent'' function $g\left(t\right) = 1-\left|t\right|$.

If $g'\left(t\right)$ diverges at one of the end points $t=\pm 1$, the supercritical regime disappears, and one of the two  phase transition is absent.  If $g'\left(t\right)$ diverges at both end points $t=\pm 1$, there are no phase transitions in the system. This is what happens for the semicircle $g\left(t\right) = \sqrt{1-t^2}$, see the Appendix.

\subsection{Generalized parabolas}
\label{Sec:parabola}

In this subsection we briefly consider a family of ``generalized parabolas'': $g\left(t\right) = 1-\left|t\right|^{\nu}$ with $\nu > 1$ (so $g$ is convex upward), thus extending the results of Sec.~\ref{Sec:parabola_nu2} where we dealt with $\nu = 2$.
Our main goal here is to understand the behavior of the system when $g\left(t\right)$ is not locally parabolic around the measurement time; hence we will only consider the measurement time $\tau = 0$.

In the subcritical regime $X  \leq  \nu$, the solution to Eq.~(\ref{eq:tangent_g}) gives the points of tangency
\begin{equation}\label{tstargen}
t_\pm= \pm \left(\frac{X-1}{\nu-1}\right)^{1/\nu} .
\end{equation}
Plugging $g\left(t\right) = 1-\left|t\right|^{\nu}$ and Eq.~(\ref{tstargen}) into Eqs.~(\ref{eq:s_t_g}) and~(\ref{eq:s_b_g}) gives the rescaled actions
\begin{eqnarray}
\tilde{s}_{\text{c}} &=& \frac{\nu^2}{2} \left(\frac{X-1}{\nu-1}\right)^{\frac{2\nu-1}{\nu}}  ,\\
\tilde{s}_{\text{u}} &=& \frac{\nu^{2}}{2\left(2\nu-1\right)}\left(\frac{X-1}{\nu-1}\right)^{\frac{2\nu-1}{\nu}}.
\end{eqnarray}
As a result,
\begin{equation}\label{ssmallgen}
s\left(X\right)=\tilde{s}_{\text{c}} -\tilde{s}_{\text{u}}=\frac{\nu^{2}\left(\nu-1\right)}{2\nu-1}\left(\frac{X-1}{\nu-1}\right)^{\frac{2\nu-1}{\nu}},\qquad 1  < X \leq \nu.
\end{equation}
In the supercritical regime, $X \ge \nu$, the optimal path $x_{\text{c}}\left(t\right)$ is given by Eq.~(\ref{eq:two_straight_lines}) (with $\tau=0$).
Evaluating the action~(\ref{eq:s_far_tail}) we find
\begin{equation}\label{slargegen}
s\left(X\right)=\frac{X^{2}}{2}-\frac{\nu^{2}}{2\left(2\nu-1\right)},\qquad X\geq\nu.
\end{equation}
As expected, the first (Gaussian) term in Eq.~(\ref{slargegen}) is universal, whereas the constant second term is contributed by the wall.

As $\tau=0$, there is no intermediate regime.  The third derivative $d^{3}S/dX^{3}$ jumps at $X=\nu$, corresponding to a third-order dynamical phase transition. The order of the transition does not depend on $\nu$ (to remind the reader, $\nu>1$ here) because the wall function $g\left(t\right)$  has the same asymptotic behavior (up to numerical coefficients) at $t=\pm 1$.

Now let us consider the near tail $X-1\ll 1$.
The analysis, which we performed in Eqs.~(\ref{eq:t_pm_g})-(\ref{eq:t_c}), is not valid for $\nu\neq 2$, because $g''\left(t=0\right)$ either vanishes or diverges. A slightly modified analysis yields the $\nu$-dependent scaling behaviors
\begin{equation}
\label{DeltaXgen}
X-1 \sim T^{\frac{\nu-1}{2\nu-1}}
\end{equation}
and
\begin{equation}\label{ellgen}
t_c \sim T^{\frac{2(\nu-1)}{2\nu-1}};
\end{equation}
The growth exponent $\beta$ and the correlation exponent $1/z$ are
\begin{equation}\label{betaznu}
\beta=\frac{\nu-1}{2\nu-1}\qquad\text{and}\qquad\frac{1}{z}=2\beta =\frac{2\left(\nu-1\right)}{2\nu-1}.
\end{equation}
{The KPZ exponents $\beta=1/3$ and $1/z=2/3$ are recovered for $\nu=2$.
As $\nu$ increases from $1$ to $\infty$,  $\beta$ increases monotonically from $0$ to $1/2$, and $1/z$ increases monotonically from $0$ to $1$. Note that, as $\nu \to \infty$,  $\tilde{s}_{\text{c}}$ tends to $\Delta X^2/2$ while $\tilde{s}_{\text{u}}$ goes to zero. In this limit $g\left(t\right)$ becomes a rectangle, and the statistical properties of $\Delta X$ coincide with those of a Brownian excursion without an absorbing wall.

Remarkably, the exponents $\beta$ and $1/z$ coincide with their counterparts for a different model:  Brownian motion in 2+1 dimensions, conditioned to stay away from a stationary absorbing wall, see Ref.~\cite{Nechaev}, Eqs.~(7) and (8).
One only needs to identify the characteristic size  $R$ of their absorbing wall with our $T$, their coordinate $x$ with our $t_c$, and their coordinate $y$ with our $X$. The authors of Ref. \cite{Nechaev} obtained the same exponents from simple scaling arguments, based on geometrical considerations. It would be interesting to investigate the origin of the coincidence of the exponents in these two models.

\subsection{Fractional-order phase transitions}
\label{sec:fractional_order}

Up to now we have seen in this system dynamical phase transitions of the third order. We now show that phase transitions of other orders are possible too. As a first example,  consider the cosine wall $g\left(t\right)=\cos\left(\pi t/2\right)$ in the case $\tau=0$.
Here the tangent lines to $g\left(t\right)$ at $t=\pm 1$ yield the critical value $X=\pi/2$.
In the subcritical regime $1 \le  X \le \pi / 2$, Eqs.~(\ref{eq:tangent_g}),~(\ref{eq:s_t_g}) and~(\ref{eq:s_b_g}) yield the large deviation function $s\left(X\right)$ in a form parametrized by $t_{-}\in\left[-1,0\right]$:
\begin{equation}
\begin{array}{ccc}
X & = & \cos\left(\frac{\pi t_{-}}{2}\right)+\frac{\pi t_{-}}{2}\sin\left(\frac{\pi t_{-}}{2}\right),\quad\\
s & = & \frac{\pi}{16}\left[\pi t_{-}\cos(\pi t_{-})-\sin(\pi t_{-})\right],
\end{array}\qquad X \le \frac{\pi}{2}.
\end{equation}
In the supercritical regime $ X \ge \pi / 2$, Eq.~(\ref{eq:s_far_tail}) leads to
\begin{equation}
s\left(X\right)=s_{\text{c}}-s_{\text{u}}=\frac{X^{2}}{2}-\frac{\pi^{2}}{16},\qquad X\ge\frac{\pi}{2}.
\end{equation}
As expected, the behavior of $s\left(X\right)$ around $X=\pi/2$ is non-analytic:
\begin{equation}
s\left(X\right)=\begin{cases}
\frac{\pi^{2}}{16}+\frac{1}{2}\pi\left(X-\frac{\pi}{2}\right)+\frac{1}{2}\left(X-\frac{\pi}{2}\right)^{2}, & X\geq \frac{\pi}{2},\\
\frac{\pi^{2}}{16}+\frac{1}{2}\pi\left(X-\frac{\pi}{2}\right)+\frac{1}{2}\left(X-\frac{\pi}{2}\right)^{2}+\frac{16}{5\pi^{3/2}}\left(\frac{\pi}{2}-X\right)^{5/2} + \dots, & \frac{\pi}{2}-X\ll1.
\end{cases}
\end{equation}
However, in contrast to the previous cases, the phase transition here at $X=\pi/2$ is of the \emph{fractional} order $5/2$ \citep{Hilfer}. What is the reason for this special behavior? Looking closely at the behavior of the cosine wall at $t= -1$, we find that it is non-generic because the quadratic term in the expansion
\begin{equation}
g\left(t\right)=\cos\left(\frac{\pi t}{2}\right)=\frac{\pi\left(t+1\right)}{2}-\frac{\pi^{3}\left(t+1\right)^{3}}{48} + \dots
\end{equation}
is absent (and similarly at $t=1$).

We now show that it is indeed the local behavior of $g\left(t\right)$ near $t=\pm1$ which determines the order of the transition. Let us consider a (convex upward) $g\left(t\right)$ whose behavior around $t = -1$ is
\begin{equation}
\label{eq:g_local_behavior_at_minus1}
g\left(t\right)=a\left(t+1\right)-b\left(t+1\right)^{n}+\dots,
\end{equation}
with $a,b > 0$ and $n > 1$, and consider arbitrary $\tau$.
A phase transition occurs along the line $X=\left(\tau+1\right)g'\left(-1\right)$ which is tangent to $g\left(t\right)$ at $t=-1$.
The nonanalytic behavior of $s$ at the transition is entirely captured by the nonanalytic behavior of the action along the optimal constrained path evaluated up to time $t=\tau$
\begin{equation}
s_{\text{c}}^{\left(1\right)}=\frac{1}{4}\int_{-1}^{\tau}\left(\frac{dx_{\text{c}}}{dt}\right)^{2}dt,
\end{equation}
because the remaining terms which contribute to $s$ are (in general) analytic at the transition point.
At supercritical $X$, the optimal constrained path is given by Eq.~(\ref{eq:two_straight_lines}), so that
\begin{equation}
s_{\text{c}}^{\left(1\right)}=\frac{X^{2}}{4\left(\tau+1\right)}, \qquad X\ge\left(\tau+1\right)g'\left(-1\right).
\end{equation}
When $X$ approaches the transition point from below, the scaling behavior of the quantities
\begin{equation}\label{quantities}
\delta X=\left(\tau+1\right)g'\left(-1\right)-X \qquad \text{and} \qquad \delta s=s_{\text{c}}^{\left(1\right)}-\frac{X^{2}}{4\left(\tau+1\right)}
\end{equation}
is the following:
\begin{equation}\label{eq:scalings_fractional_order}
\delta X \sim \left(t_{-}+1\right)^{n-1} \qquad \text{and} \qquad  \delta s \sim \left(t_{-}+1\right)^{2n-1},
\end{equation}
leading to
\begin{equation}
\label{eq:order_of_fractional_order_transition}
\delta s\sim\delta X^{\frac{2n-1}{n-1}},
\end{equation}
that is, the phase transition is of order $(2n-1)/(n-1)$.
The generic third-order transition (see for instance Sec.~\ref{Sec:parabola}) and the $5/2$-order transition for $g\left(t\right)=\cos\left(\pi t/2\right)$ are particular cases of Eq.~(\ref{eq:order_of_fractional_order_transition}) with $n=2$ and $n=3$, respectively. The analogy with phase transitions is enhanced by the fact that  $t_{-} \! + \! 1$ is a natural order parameter: it vanishes above the transition, but is nonzero below the transition. The scaling behaviors~(\ref{eq:scalings_fractional_order}) yield nontrivial exponents which describe the critical behavior of the system near the transition.

The same arguments apply to the other phase transition line, $X =  \left(\tau-1\right)g'\left(1\right)$, which is the tangent to $g\left(t\right)$ at $t=1$. The order of the corresponding phase transition depends on the local behavior of $g\left(t\right)$ near $t=1$.

\subsection{Tent}
\label{sec:tent}

In most of our derivations so far we assumed that $g\left(t\right)$ is smooth and strictly convex upward. What happens if the wall function $g\left(t\right)$ has a corner singularity?
It is known that, if a corner singularity coincides with the observation time $t=\tau$, the scaling of typical fluctuations is strongly affected \citep{FS}.
Here we show that the critical behavior of the system also changes: the transition becomes of the second order. A simple example is provided by the ``tent'' function $g\left(t\right) = 1-\left|t\right|$.

\begin{figure}
\includegraphics[width=0.4\textwidth,clip=]{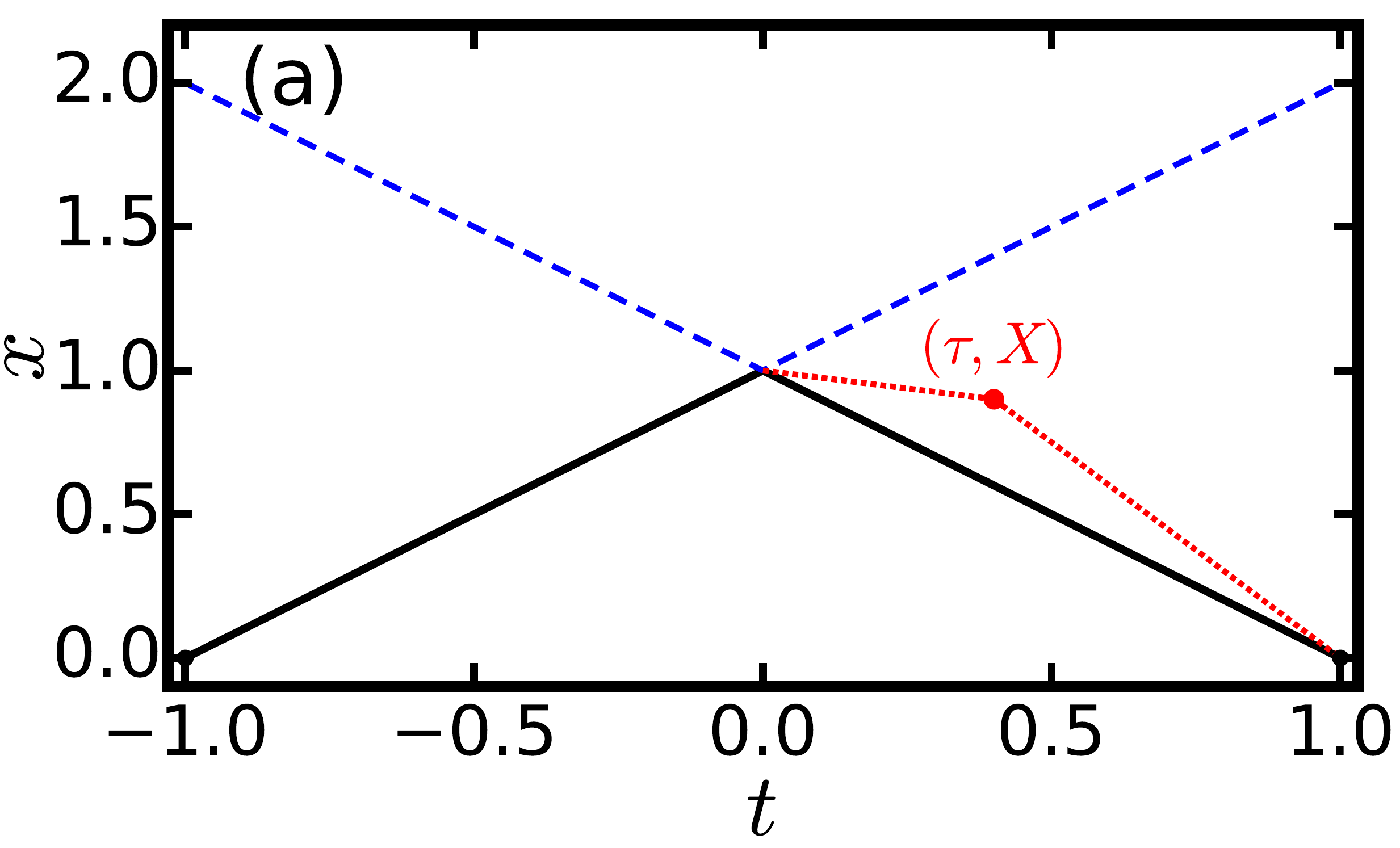}
\includegraphics[width=0.4\textwidth,clip=]{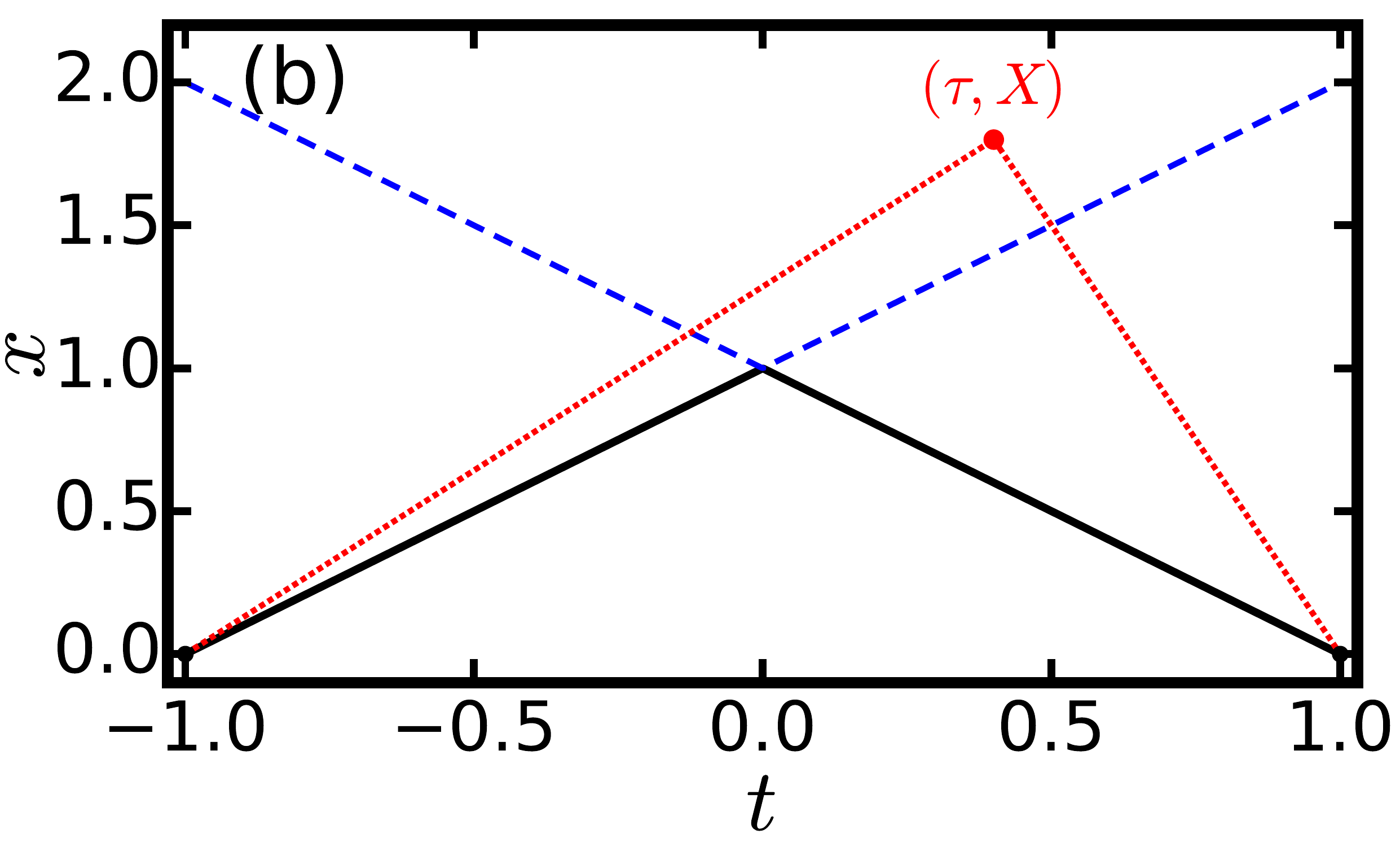}
\caption{Solid line: the wall function $g\left(t\right)=1-\left|t\right|$.
Dotted line: the optimal path constrained on $x\left(t=\tau\right)=X$, at times where $x\left(t\right)\ne g\left(t\right)$, in the subcritical (a), and supercritical (b) regimes.
A second-order phase transition occurs along the dashed line $x = 1+\left|t\right|$.}
\label{tent_domain}
\end{figure}

Since $g\left(t\right)$ is now (weakly) convex upward, the optimal unconstrained path still follows the wall, $x_{\text{u}}\left(t\right)=g\left(t\right)=1-\left|t\right|$. The action~(\ref{action2}), evaluated along this path, is $s_{\text{u}} = 1/2$.
Regarding the optimal constrained path $x_{\text{c}}\left(t\right)$, the tangent construction described in Sec.~\ref{sec:convexg_general} is not applicable because $g''\left(t\right)$ either vanishes or does not exist.
There are two regimes of interest, see Fig.~\ref{tent_domain}. In the subcritical regime $X \le 1+\left|\tau\right|$ we obtain for $\tau > 0$:
\begin{equation}
\label{eq:x_t_tent_subcritical}
x_{\text{c}}\left(t\right)=\begin{cases}
1+t, & -1\le t\le0,\\
1+\frac{t\left(X-1\right)}{\tau}, & 0\le t\le\tau,\\
\frac{\left(1-t\right)X}{1-\tau}, & \tau\le t\le1.
\end{cases}
\end{equation}
[For $\tau < 0$ the optimal path is the mirror image of Eq.~(\ref{eq:x_t_tent_subcritical}).] That is, the tangent of Sec.~\ref{sec:convexg_general} is replaced by a straight line which connects the point
$\left(\tau,X\right)$ with the point $\left(t=0,x=1\right)$ of the corner singularity of $g\left(t\right)$.
The action~(\ref{action2}), evaluated along $x_{\text{c}}\left(t\right)$, is
\begin{equation}
s_{\text{c}} = \frac{\left(1-X\right)^{2}+\left|\tau\right|\left(2X-\left|\tau\right|\right)}{4\left|\tau\right|\left(1-\left|\tau\right|\right)},
\end{equation}
yielding the large deviation function in the subcritical regime:
\begin{equation}
\label{eq:s_corner_subcritical}
s\left(X,\tau\right)=s_{\text{c}}-s_{\text{u}}=\frac{\left(1-X\right)^{2}+\left|\tau\right|\left(2X+\left|\tau\right|-2\right)}{4\left|\tau\right|\left(1-\left|\tau\right|\right)}, \qquad X\le 1+\left|\tau\right|.
\end{equation}
In the supercritical regime $X \ge 1+\left|\tau\right|$ the optimal path is given by Eq.~(\ref{eq:two_straight_lines}), and $s$ is found from Eq.~(\ref{eq:s_far_tail}) to be
\begin{equation}
\label{eq:s_corner_supercritical}
s\left(X,\tau\right)=s_{\text{c}}-s_{\text{u}}=\frac{X^{2}+\tau^{2}-1}{2\left(1-\tau^{2}\right)}, \qquad X \ge 1+\left|\tau\right|.
\end{equation}
From Eqs.~(\ref{eq:s_corner_subcritical}) and~(\ref{eq:s_corner_supercritical}) we find that it is the second derivative $\partial^{2}s/\partial x^{2}$ which jumps
along the transition line $X=1+\left|\tau\right|$. That is, the dynamical phase transition is of the second order.
One way to understand this result is to think of the tent as the $n \to \infty$ limit of the class of functions $g\left(t\right)$ from Eq.~(\ref{eq:g_local_behavior_at_minus1}). That the transition is of the second order for the tent then corresponds to the $n\to\infty$ limit of Eq.~(\ref{eq:order_of_fractional_order_transition}).

In the particular case $\tau = 0$ there is no subcritical regime and therefore no phase transition. Here the large-deviation function
\begin{equation}
\label{eq:s_traingle_tau_zero}
s\left(X, \tau =0 \right)=\frac{X^{2}-1}{2}
\end{equation}
describes a distribution which has the form of a Gaussian tail. In the near tail, $X-1 \ll 1$, Eq.~(\ref{eq:s_traingle_tau_zero}) yields
\begin{equation}
\label{eq:P_traingle_tau_zero}
-\ln \mathcal{P}\simeq\frac{C\left(X-CT\right)}{D},\qquad X-CT\ll CT,
\end{equation}
predicting a $T$-independent scaling of typical fluctuations of $X$. This result can be also obtained by taking the limit $\nu \to 1$ in Eq.~(\ref{DeltaXgen}). Also, this result is in agreement with the corresponding result quoted in Ref. \cite{FS}%
\footnote{Plugging the parameters $C=1$, and $D=1/2$ of Ref. \cite{FS} into Eq.~(\ref{eq:P_traingle_tau_zero}), we obtain $\mathcal{P}\sim e^{-2\left(X-T\right)}$, which agrees with the result obtained in Sec. 5 (i.a) of Ref. \cite{FS} with $g\left(t\right) = 1-\left|t\right|$ (up to a preexponential factor which is beyond the accuracy of our leading-order OFM approximation).}.

\section{Non-convex $x_{0}\left(t\right)$}
\label{sec:nonconvex}

The OFM formulation of Sec.~\ref{ConstrainedBrownianExcursion} is valid regardless of the convexity of $g\left(t\right)$. However, if $g\left(t\right)$ is not convex upward, finding the optimal paths can become more involved technically. Still, we can make some general observations.

Importantly, for non-convex $g\left(t\right)$ the optimal unconstrained path $x_{\text{u}}\left(t\right)$ does \emph{not} coincide with $g\left(t\right)$, but rather with its convex envelope $g_{c}\left(t\right)$, see Fig.~\ref{nonconvex_wall}. As a result, the peak of the distribution of $X$, at given $\tau$, is around $X=g_{c}\left(\tau\right)$.  This is in contrast to the convex-upward case, where the distribution is peaked at a point which is much closer to the wall.
In the regime where $g\left(\tau\right)\ne g_{c}\left(\tau\right)$, typical fluctuations of $X$ around $g_{c}\left(\tau\right)$ are normally distributed, and this Gaussian asymptotic of the distribution can be described by the OFM. Moreover, now the complete distribution has another tail, where $X - g\left(\tau\right)$ is \emph{negative} and much larger in absolute value than its typical value. This tail can also be obtained using the OFM.

At $X \ge g_{c}\left(\tau\right)$ the large-deviation function $s\left(X,\tau\right)$ depends on $g\left(t\right)$ only through its convex envelope $g_{c}\left(\tau\right)$.
A similar observation for typical fluctuations in the regime where $g\left(\tau\right)\ne g_{c}\left(\tau\right)$ was made in Ref.~\cite{FS}, Sec.~5(iii).
As a result, dynamical phase transitions occur along lines in the $tx$ plane which are tangent to $g_{c}\left(t\right)$ at $t = \pm 1$ where these lines do not coincide with $g_c$ itself, see Fig.~\ref{nonconvex_wall}.

\begin{figure}
\includegraphics[width=0.4\textwidth,clip=]{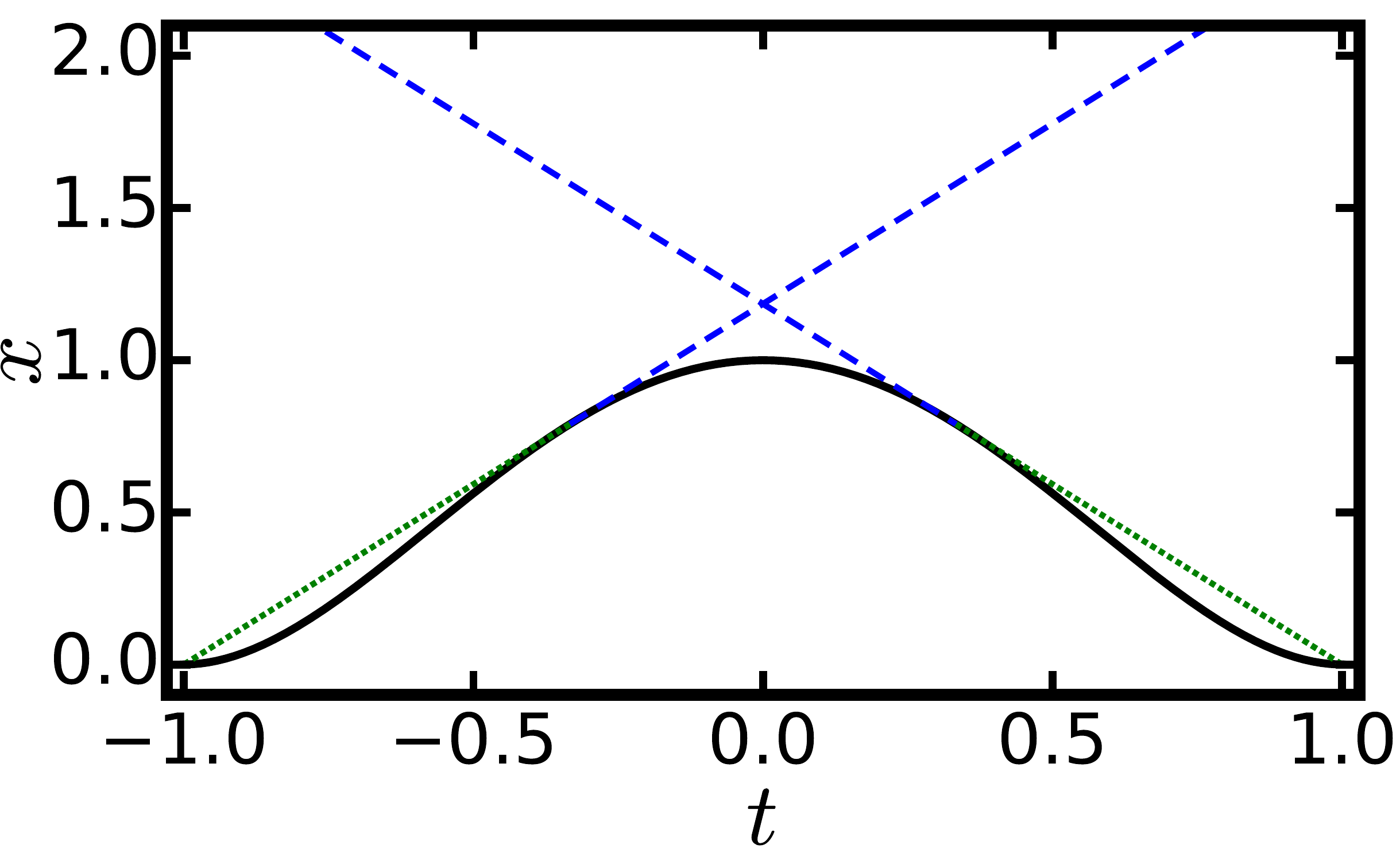}
\caption{Solid line: the wall function $g\left(t\right)=(1-t^{2})^2$. The optimal unconstrained path $x_{\text{u}}\left(t\right)$ (the dotted line) coincides with $g_c\left(t\right)$, the convex envelope of $g\left(t\right)$.
Dashed line: the continuation of the tangents of $g_c\left(t\right)$ at $t = \pm 1$. The latter line signifies a dynamical phase transition.}
\label{nonconvex_wall}
\end{figure}

\section{Summary and discussion}
\label{disc}

We studied the distribution $\mathcal{P}\left(X,\tau,T\right)$ of the position $X = x\left(t=\tau\right)$ of a Brownian excursion $x\left(t\right)$ conditioned on staying away from a moving wall $x_0\left(t\right)$ \citep{Groeneboom1989, Frachebourg2000, FS}. We focused on large deviations of $X$ and calculated the corresponding large deviation function $s\left(X,\tau\right)$ by using the optimal fluctuation method (OFM), which in this context coincides with geometrical optics. The ensuing standard variational problem can be solved by means of a simple geometric construction.
Despite the simplicity of the model, its behavior is quite rich.
The OFM correctly describes the near tail of the distribution and therefore captures the scaling behavior of typical fluctuations of $X$.
The system exhibits dynamical phase transitions -- singularities of the large deviation $s\left(X,\tau\right)$ -- for a broad class of wall functions $x_{0}\left(t\right)$. The transitions occur due to a qualitative change in the character of the optimal path as $X$ and/or $\tau$ are changed.

Until now, many instances of dynamical phase transitions -- that is, singularities of large-deviation functions -- have been observed in one-particle and multi-particle systems \citep{Graham,Jauslin,DMS,Schuetz,Derrida2007,bertini2011,Lecomte,shortreview,hurtadoreview, Baek2015,Janas2016,Touchette, Baek2017,Baek2018, SKM2018}. Many of them have been  described by the OFM \citep{Graham,Jauslin,DMS,bertini2011,Lecomte,hurtadoreview, Baek2015,Janas2016,Baek2017,Baek2018, SKM2018}. In the OFM description, the singularities are usually caused either by a switching between two different optimal paths at the critical point (for first order transitions), or by a spontaneous symmetry breaking of the optimal path (for second order transitions). In contrast, the mechanism which causes the phase transitions of the constrained Brownian excursion is geometrical by its nature and is analogous to shadows in optics. The transition occurs when the observation point $\left(\tau,X\right)$ enters a complete or partial space-time ``shadow'' of the wall $x_{0}\left(t\right)$.
Remarkably, this simple mechanism can lead to different orders of the transition. For a generic convex upward $x_0\left(t\right)$, it is of the third order. However, fractional orders are also possible. We showed that the order of transition depends on the local behaviors of $x_0\left(t\right)$ at $t = \pm T$.
Second order transitions are possible if $x_0\left(t\right)$ has a corner singularity at some time.

Recently third-order transitions have been discovered in large deviation functions of several stochastic many-body systems, see  Ref. \cite{shortreview}  for a concise review. These systems include Gaussian random matrices, nonequilibrium stochastic growth models  belonging to the KPZ  universality class
and, tantalizingly, $N\gg 1$ non-intersecting Brownian excursions in $1+1$ dimension \cite{shortreview,3orderKPZ}. It may be tempting to lump together all these third order transitions. There are, however, important differences between the FS model and the other models mentioned above. First, in the FS model the
phase transition point is located outside the region of typical fluctuations. Second, the large-deviation function of the FS model has the same scaling behavior, as a function of $T$, below and above the transition. Third, the typical fluctuations of the FS model are described by the FS distribution \cite{FS}, rather than the Tracy-Widom distribution \cite{TW}. These differences and the simple geometric mechanism, which is present in the FS model transition and apparently absent in the other models, show that the third-order transition in the FS model has a different nature.

For a wall function $x_0\left(t\right)$ which is not convex upward, we found that the peak of the distribution $\mathcal{P}\left(X,\tau,T\right)$ is near the convex envelope of $x_0\left(t\right)$. At times $\tau$ where $x_0\left(\tau\right)$ is not equal to its convex envelope, typical fluctuations around this peak follow a Gaussian distribution which can be calculated with the OFM.

Somewhat counter-intuitively, the leading-order OFM approximation, which we used here, yields the same large-deviation function $s\left(X,\tau\right)$ for absorbing and reflecting walls. So all of our large-deviation results can be immediately extended to reflecting walls. The difference between absorbing and reflecting walls should be very pronounced in the region of typical, small fluctuations which are beyond the OFM validity. The difference should also appear in the pre-exponential factors that we did not  calculate.

Finally, there is a fascinating connection between the geometrical optics of constrained Brownian motion and the recently suggested Tangent Method of determining
the so called Arctic curve \cite{Colomo}. The Arctic curve is the boundary between``frozen" and ``liquid" regions in several two-dimensional discrete models of statistical mechanics which exhibit phase-separation because of  finite-size effects.
As shown in Ref. \cite{Colomo}, basic excitations in these models ``form random walks from a given boundary point to the Arctic curve, which are almost straight in the thermodynamic limit, and reach
the curve tangentially." Not surprisingly, the Tangent Method has considerably simplified the calculations of Arctic curves \cite{FL2018,FG2018}.

\section*{Acknowledgments}

We acknowledge useful discussions with Tal Agranov, Mark Dykman, Patrik Ferrari and Senya Shlosman. This research was supported by the Israel Science Foundation (grant No. 807/16). N.R.S. was supported by the Clore Foundation.


\section*{Appendix: semicircle $g\left(t\right) = \sqrt{1-t^2}$}

\renewcommand{\theequation}{A\arabic{equation}}
\setcounter{equation}{0}

Here we consider the particular case of a semicircle $g\left(t\right)=\sqrt{1-t^{2}}$ in some detail.
The solutions to Eq.~(\ref{eq:tangent_g}) are
\begin{equation}\label{t1andt2}
t_{\pm}=\frac{\tau \pm X \sqrt{\tau ^2+X^2-1}}{\tau ^2+X^2}.
\end{equation}
We plug $g\left(t\right)=\sqrt{1-t^{2}}$ and Eq.~(\ref{t1andt2}) into Eqs.~(\ref{eq:s_t_g}) and~(\ref{eq:s_b_g}) to obtain
\begin{equation}
\tilde{s}_{\text{c}}= \frac{X \sqrt{\tau ^2+X^2-1} \left(2 \tau ^2+X^2-1\right)}{2
   \left(1-\tau ^2\right) \left(\tau ^2+X^2\right)}
\end{equation}
and
\begin{equation}
\tilde{s}_{\text{u}} = \frac{1}{4}\int_{t_-}^{t_+} \frac{t^2}{1-t^2}\,dt =\frac{1}{4}\, \text{arctanh}\left( \frac{2 X \sqrt{\tau ^2+X^2-1}}{\tau^2+2 X^2-1}\right)  -\frac{X \sqrt{\tau ^2+X^2-1}}{2 \left(\tau^2+X^2\right)} .
\end{equation}
\begin{figure}
\includegraphics[width=0.4\textwidth,clip=]{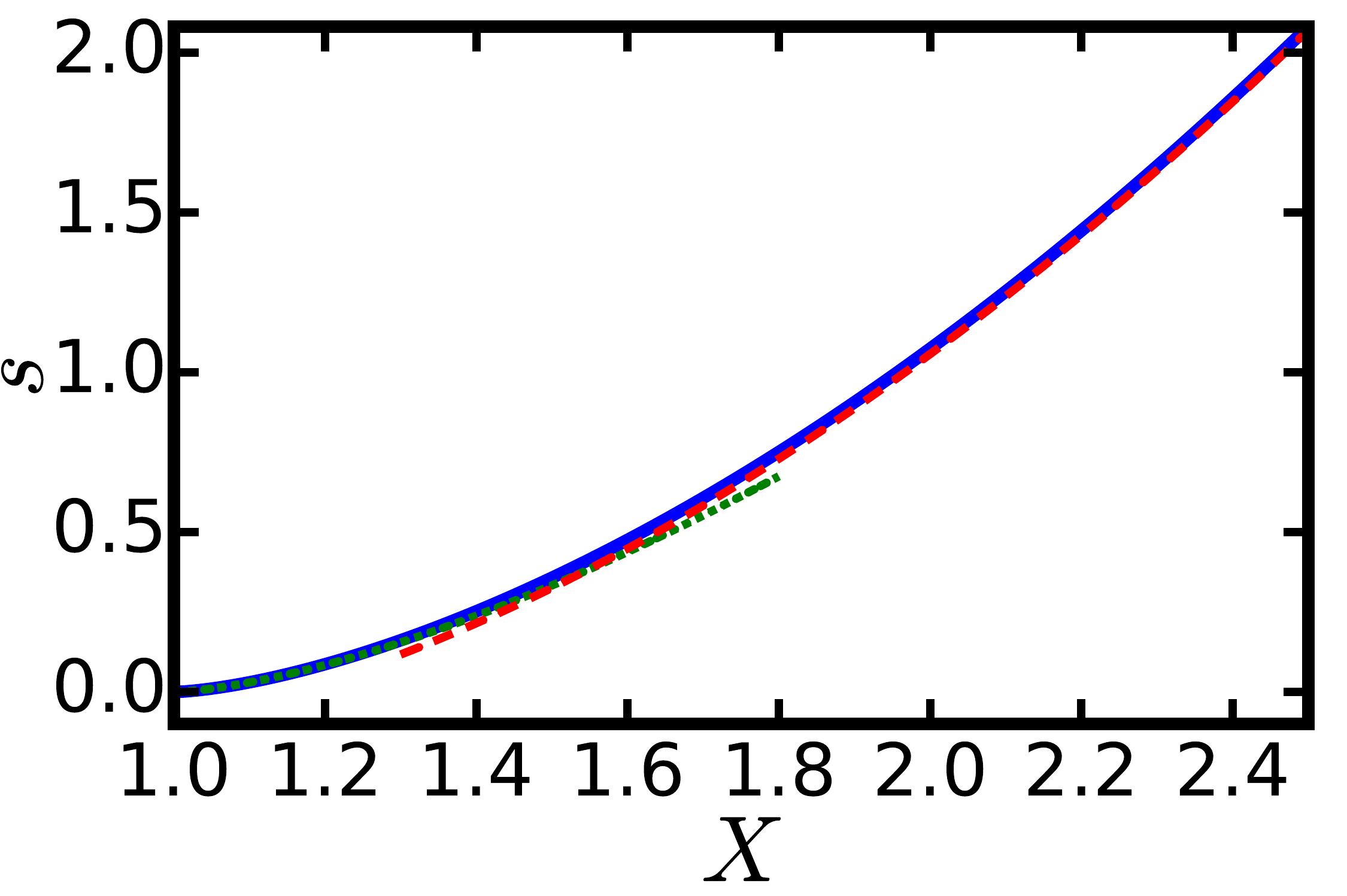}
\caption{The large-deviation function $s\left(X,\tau=0\right)$ for the semicircle $g\left(t\right)=\sqrt{1-t^{2}}$ (solid), see Eq.~(\ref{ssmalcirc}), together with its near- and far-tail asymptotics~(\ref{TWtailcir}) and~(\ref{eq:circFar}) (dotted and dashed, respectively).}
\label{s_X_semicircle}
\end{figure}
As a result, the large deviation function $s\left(X,\tau\right)$, for all $X>g(\tau)$, is the following:
\begin{equation}
\label{ssmalcirc}
s\left(X,\tau\right)=\tilde{s}_{\text{c}}-\tilde{s}_{\text{u}} =\frac{X\sqrt{\tau^{2}+X^{2}-1}}{2\left(1-\tau^{2}\right)}-\frac{1}{4}\text{arctanh}\left(\frac{2X\sqrt{\tau^{2}+X^{2}-1}}{\tau^{2}+2X^{2}-1}\right).
\end{equation}
Together with Eq.~(\ref{action1}), Eq.~(\ref{ssmalcirc}) gives, up to pre-exponential factors, the probability distribution $\mathcal{P}\left(X,\tau,T\right)$ in the original variables.
The near-tail asymptotic
of Eq.~(\ref{ssmalcirc}) is
\begin{equation}\label{TWtailcir}
s\left(X,\tau\right)\simeq\frac{2\sqrt{2}\left(\Delta X\right)^{3/2}}{3\left(1-\tau^{2}\right)^{3/4}},\qquad \text{where}\quad \Delta X = X-\sqrt{1-\tau^2}\ll \sqrt{1-\tau^2}.
\end{equation}
The same result follows from Eq.~(\ref{eq:s_typical_fluctuations_convex_g}) with $g\left(t\right) = \sqrt{1 - t^2}$.
The plot of $s\left(X,\tau=0\right)$ is shown in Fig.~\ref{s_X_semicircle}, alongside with the near tail asymptotic (\ref{TWtailcir}) and
the far tail asymptotic
\begin{equation}
\label{eq:circFar}
s\left(X,\tau\right)=\frac{X^2}{2 \left(1-\tau ^2\right)}-\frac{1}{4} \ln
   \left(\frac{X^2}{1-\tau ^2}\right)-\frac{1}{4}-\frac{\ln2}{2}+\dots, \qquad X\gg \sqrt{1-\tau^2}.
\end{equation}

Now let us return to the near-tail asymtotic (\ref{TWtailcir}) and
plug it into~(\ref{action1}) with $\gamma=1$. We obtain
\begin{equation}
\label{TWtailcirc1}
-\ln\mathcal{P}\simeq\frac{2\sqrt{2C}\left(X-C\sqrt{T^{2}-\tau^{2}}\right)^{3/2}}{3D\left(1-\frac{\tau^{2}}{T^{2}}\right)^{3/4}T^{1/2}},\qquad X-C\sqrt{T^{2}-\tau^{2}}\ll C\sqrt{T^{2}-\tau^{2}},
\end{equation}
in the original variables. We now show that this result agrees with the \emph{tail} of the Ferrari-Spohn (FS) distribution \cite{FS} of typical fluctuations of $X$ away from the circle\footnote{see footnote \ref{footnote:FScomment}.}.
FS introduced a stationary diffusion process $\mathcal{A}(t)$, described by the Langevin equation
\begin{equation}
\frac{d\mathcal{A}}{dt}=a(\mathcal{A})+\xi\left(t\right),
\end{equation}
with the Gaussian white noise $\xi\left(t\right)$, as described by Eq.~(\ref{whitenoise}) with $D=1/2$, and the drift term
\begin{equation}\label{drift}
a\left(\mathcal{A}\right)=\frac{\text{Ai}'\left(\mathcal{A}-\omega_{1}\right)}{\text{Ai}\left(\mathcal{A}-\omega_{1}\right)}.
\end{equation}
Here $\text{Ai}(\dots)$ is the Airy function, and $-\omega_1=-2.338107\dots$ is its first zero.
The equilibrium probability distribution of this process,
\begin{equation}\label{FSdist}
\rho\left[\mathcal{A}\left(t\right)=z\right]=\frac{\left[\text{Ai}\left(z-\omega_{1}\right)\right]^{2}}{\left[\text{Ai}'\left(-\omega_{1}\right)\right]^{2}},
\end{equation}
is depicted in Fig.~\ref{rho(z)}.
\begin{figure}
\includegraphics[width=0.4\textwidth,clip=]{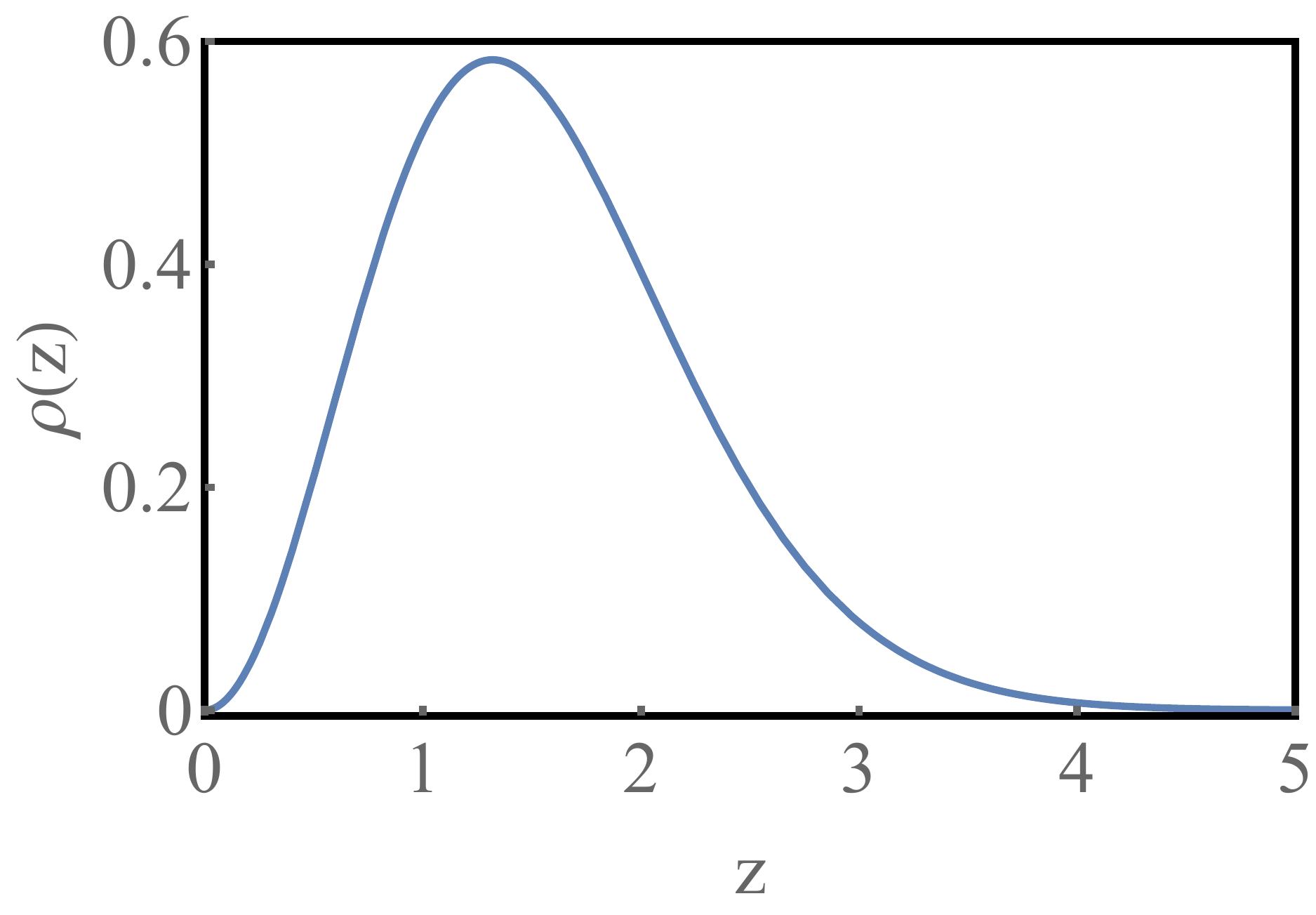}
\caption{The equilibrium probability distribution of the Ferrari-Spohn process, Eq.~(\ref{FSdist}).}
\label{rho(z)}
\end{figure}
As FS proved, at $T \to \infty$, typical fluctuations of $\Delta X\!=\!X-C\sqrt{T^{2}-\tau^{2}}$ away from the circle are distributed as $\mathcal{A}\left(2/T\right)^{-1/3}\left(1-\tau^{2}/T^{2}\right)^{1/2}$.
That is, in terms of $\Delta X$ the distribution is
\begin{equation}\label{PFS}
\mathcal{P}\left(\Delta X,\tau,T\right)=\frac{\left(2/T\right)^{1/3}\left\{ \text{Ai}\left[\frac{\left(2/T\right)^{1/3}\Delta X}{\left(1-\frac{\tau^{2}}{T^{2}}\right)^{1/2}}-\omega_{1}\right]\right\} ^{2}}{\left(1-\frac{\tau^{2}}{T^{2}}\right)^{1/2}\,\left[\text{Ai}'\left(-\omega_{1}\right)\right]^{2}} .
\end{equation}
The tail of this distribution is given by the large-argument asymptotic of the Airy function:
\begin{equation}\label{tailFS}
\mathcal{P}\left(\Delta X\to\infty,\tau,T\right)=\frac{\exp\left[-\frac{4\sqrt{2}\,\left(\Delta X\right)^{3/2}}{3\left(1-\frac{\tau^{2}}{T^{2}}\right)^{3/4}T^{1/2}}\right]}{2^{11/6}T^{1/6}\pi\left[\text{Ai}'\left(-\omega_{1}\right)\right]^{2}\left(\Delta X\right)^{1/2}\left(1-\frac{\tau^{2}}{T^{2}}\right)^{1/4}}.
\end{equation}
Our near-tail result (\ref{TWtailcirc1}) coincides with this asymptotic up to the pre-exponential factor, which is unaccounted for by the leading-order OFM.
This comparison confirms that the OFM is valid for large deviations, $-\ln\mathcal{P} \gg 1$, starting from the near tail and toward larger $X$.

\end{document}